\documentclass[prd,twocolumn]{revtex4}
\usepackage{amssymb,amsmath,bm,dsfont}

\newcommand\deli{\Delta}
\newcommand\delii{\Delta_\Gamma}
\newcommand\intx{\int\!d^3{\bm x}}
\newcommand\hdual{{\bm \theta}}
\newcommand\hmean{\bar h}
\newcommand\hhat{\hat{\bm h}}
\newcommand\hdualhat{\hat{\bm \theta}}
\newcommand\mc{m_{\rm c}}
\newcommand\mk{m_{\rm k}}
\newcommand\mck{m_{\rm c,k}}
\newcommand\tc{t_{\rm c}}
\newcommand\tk{t_{\rm k}}
\newcommand\tck{t_{\rm c,k}}

\begin{document}

\title{Evolution of fluctuations near QCD critical point}

\author{M.~A.~Stephanov}
\affiliation{Department of Physics, University of Illinois, Chicago, 
Illinois 60607, USA}

\date{November 2009}

\begin{abstract}
We propose to describe the time evolution of quasi-stationary
fluctuations near QCD critical point by a system of stochastic
Boltzmann-Langevin-Vlasov-type equations. We derive the equations and
study the system analytically in the linearized regime.  Known results
for equilibrium stationary fluctuations as well as the critical
scaling of diffusion coefficient are reproduced. We apply the approach
to the long-standing question of the fate of the critical point
fluctuations during the hadronic rescattering stage of the heavy-ion
collision after chemical freezeout. We find that if conserved particle
number fluctuations survive the rescattering, so do, under a certain
additional condition, the fluctuations of non-conserved quantities,
such as mean transverse momentum. We derive a simple analytical
formula for the magnitude of this ``memory'' effect.
\end{abstract}


\maketitle

\section{Introduction}
\label{sec:intro}

Mapping the QCD phase diagram as a function of temperature $T$ and
baryochemical potential $\mu_B$ is one of the fundamental goals of
heavy-ion collision experiments. QCD critical point is a distinct
singular feature of the phase diagram. It is a ubiquitous property of
QCD models based on the chiral symmetry breaking
dynamics~\cite{Asakawa:1989bq,Barducci:1989wi} (see
\cite{Stephanov:2004wx} for review and further references). Locating
the point using first-principle lattice calculations is a formidable
challenge~\cite{Fodor:2001pe,Ejiri:2003dc,Gavai:2004sd,Gavai:2008zr,deForcrand:2003bz}.
Recent progress and results are encouraging, but much work needs to be
done to understand and constrain systematic errors (see,
e.g., Refs.\cite{Schmidt:2009qq,Gupta:2009mu,Philipsen:2009dn} and
reviews~\cite{Schmidt:2006us,Stephanov:2007fk} for further
references and discussion).  

If the critical point is located in the
region accessible to heavy-ion collision experiments it can be
discovered experimentally. The search for the critical point is
planned at the Relativistic Heavy Ion Collider (RHIC) at BNL, the
Super Proton Synchrotron (SPS) at CERN, the future Facility for
Antiproton and Ion Research (FAIR) at GSI, and Nuclotron-based Ion
Collider Facility (NICA) in
Dubna~\cite{Mohanty:2009vb2,Schuster:2009ak,Stefanek:2009,CPOD}.

The characteristic feature of a critical point is the increase and
divergence of fluctuations. The non-monotonous behavior of
event-by-event fluctuations, measured in heavy-ion
collisions, as a function of the initial collision energy is a
signature of the QCD critical
point~\cite{Stephanov:1998dy,Stephanov:1999zu}. 
The estimates of the magnitude of the fluctuations in
\cite{Stephanov:1999zu} were based on the assumption of thermodynamic
equilibrium, which is a reasonable first approximation at
freezeout. For such stationary fluctuations the probability of a given
value of a fluctuating variable is proportional to the exponential
of the entropy, i.e., to the number of microscopic states with that
value of the variable~\cite{Einstein:1910,Lifshitz:v5}.

In a dynamic environment of a heavy ion collision, the system
continuously evolves with time. As long as the evolution is slow
enough compared to the typical re-equilibration time, one can consider
fluctuations as simply tracking the evolving equilibrium
conditions. However, some fluctuating modes can be slower. In fact, it
is precisely these slow modes which are of primary interest to
us. These include fluctuations of conserved quantities and, most
importantly, the critical fluctuations of the order parameter field
$\sigma$ at the critical point. Fluctuations must keep readjusting to
the continuously drifting equilibrium value. Can this quasi-stationary
dynamics of fluctuations be described quantitatively? The purpose of this
paper is to achieve this.

We derive stochastic equations for the particle distribution functions
as well as the critical mode using fluctuation-dissipation relation in
Section~\ref{sec:formalism}. We determine the corresponding equation
for the correlators of the fluctuations in
Section~\ref{sec:time-evol-fluc} and study its solution in
Sections~\ref{sec:const-coeff}, ~\ref{sec:slowest-mode}
and~\ref{sec:non-zero-modes}. Finally, in
Section~\ref{sec:chem-to-kin}, as an example of the
application, we answer analytically, in an idealized regime, 
the long-standing question of the fate of
fluctuations during the hadronic rescattering phase. We discover
a ``memory'' effect, which protects not only fluctuations of {\em
  conserved} quantities. Notations introduced throughout the paper are
indexed in Appendix~\ref{sec:notations}.

\section{Comparison with related work}
\label{sec:comp-with-relat}

The time evolution of fluctuations  has been considered previously
using different methods and/or in different contexts. Below we
review some of this work in order to point out the new ingredients as
well as the results of our approach.

Quasi-stationary dynamics of fluctuations
motivated Ref.~\cite{Berdnikov:1999ph}. The relaxation of the
correlation length, as a proxy to the magnitude of fluctuations, was
studied using a model equation. Here we shall
address evolution of fluctuations on a
microscopic level and directly in terms of observable quantities.

The quasi-stationary dynamics plays an essential role in the anomalous
suppression of charge fluctuations, which has been proposed as a
signature of the quark-gluon plasma formation at early times in
Refs.~\cite{Jeon:2000wg,Asakawa:2000wh}, and more quantitatively
analyzed in \cite{Shuryak:2000pd} (see also review
Ref.~\cite{Koch:2008ia}). The evolution follows a diffusion-type
equation, which means that fluctuations of larger spatial extent relax
slower. Therefore, as the size of the acceptance window is increased,
the ``memory'' of the fluctuations goes further back in time, allowing
to probe earlier stages of the fireball evolution.

The most interesting and not easily anticipated result of the
approach we introduce here is the following. Although the ``memory''
effect is due to the slowness of the {\em conserved} charge
fluctuations, the fluctuations of other quantities are also
affected. For example, we show that the 
fluctuations of observables such as, e.g., mean transverse momentum
$p_T$ in the event also ``remember'' their earlier value. More
precisely, if the chemical freezeout (the freezeout of inelastic
reactions) has occurred near the critical point, the elastic
collisions during the subsequent evolution of the fireball do not
completely ``wash out'' the critical point contribution to the mean
$p_T$ fluctuations even on the time scales longer than typical collisional
relaxation time. We can determine the magnitude of that effect by
studying the microscopic nature and evolution of the slowest mode of
fluctuations.

The evolution of fluctuations in the vicinity of the critical point
has been studied numerically in Ref.~\cite{Paech:2005cx}.  
The fluctuations 
were introduced by randomization of initial conditions, while the
subsequent evolution was deterministic. The essential ingredient of
the approach we introduce here is the full treatment of
fluctuations. I.e., fluctuations are driven by a random external
source, acting at all times. The strength of the source is determined
by fluctuation-dissipation relation. 

Stochastic Langevin-type equations were used to study hydrodynamic fluctuations near the critical point in~\cite{Son:2004iv}. 
However, experimental
 observables such as, e.g., mean $p_T$ fluctuations,
 are not directly related to hydrodynamic variables. The
new ingredient in the present approach is the use of kinetic Boltzmann
equation. The degrees of
freedom here are particle distribution
 functions, which directly translate into observable fluctuation
 measures. Strictly speaking, our approach lacks rigorous consistency
 of low-energy hydrodynamic description, and should be considered as
 only a model of the late hadronic stage of the heavy-ion
 collision. However, this relatively minor compromise allows us to
 address directly experimental fluctuation measures
 and make quantitative predictions, rather than limiting the
study to density fluctuations. As a test of the new approach we
shall derive some results of Ref.~\cite{Son:2004iv} in
Section~\ref{sec:hydro-mode}.

In a different context, the stochastic Boltzmann-Vlasov type
equations have been used to estimate the rate for hot electroweak baryon
number violation~\cite{Huet:1996sh,Bodeker:1998hm,Arnold:1998cy}.


\section{The formalism}
\label{sec:formalism}

To model the fireball evolving through the phase diagram near the
critical point, we consider a relativistic system of particles
interacting with a scalar field $\sigma$, coupled to a thermal bath at
temperature $T$. We are considering late hadronic phase of the
fireball expansion. The particle density is assumed to be already
sufficiently small, so that their motion can be considered
classically, using Boltzmann equation. The mass $m$ of the scalar
field, on the other hand, is considered to be sufficiently small
compared to $2\pi T$, so that the field can be treated classically,
using field equations. This condition is fulfilled sufficiently close
to the critical point.

\subsection{Equations of motion}
\label{sec:eq-motion}

Since the field $\sigma$ is a Lorentz scalar, we assume that the
coupling of it to the particles affects their mass (as opposed to,
e.g., chemical potential, which is a Lorentz vector). As an example,
one can consider coupling of nucleons $\sigma\bar N N$ or pions
$\sigma\pi\pi$ to the $\sigma$ field in the chiral sigma model. Thus we
are led to consider classical motion of particles with variable mass
$M(\sigma)$~\cite{BarrabesHenry:1976}, which depends on the local
value of the scalar field $\sigma$. The action of the system is given
by:
\begin{equation}
  \label{eq:action}
  {\cal S} 
= \intx\frac12\big(\partial_\mu\sigma\partial^\mu\sigma - U(\sigma)\big) -
\int\! ds\, M(\sigma),
\end{equation}
where the last integral is taken over the worldline of a particle with
variable mass. The corresponding equations of motion are given by:
\begin{equation}
  \label{eq:sigma-eom}
  \partial^2\sigma + dU/d\sigma + \int\!ds\,(dM/d\sigma)=0;
\end{equation}
\begin{equation}
  \label{eq:particle-eom}
  dp^\mu/d\tau = \partial^\mu M(\sigma),
\quad 
\textrm{with}\quad p^\mu=M dx^{\mu}/d\tau,
\end{equation}
where $M=M(\sigma)$ is the local value of the variable particle mass.
One can check that the motion governed by equations
(\ref{eq:particle-eom}) preserves
$p^\mu p_\mu - M(\sigma)^2=0$ along the particle trajectory.

The Boltzmann equation for the distribution function $f(x,p)$ of such
particles in the external field $\sigma$ reads~\cite{Stewart:1971}:
\begin{equation}
  \label{eq:boltzmann-part}
  \frac{p^\mu}{M}\,\frac{\partial f}{\partial x^\mu} 
+  \partial^\mu M\,\frac{\partial f}{\partial p^\mu} + {\cal C}[f] =0\,,
\end{equation}
or in a more physically transparent, non-covariant, form:

\begin{equation}
  \label{eq:boltz-noncov}
  \dot {f} +  \bm {v \cdot \nabla} f 
- (\bm\nabla M/\gamma)\bm\cdot (\partial f/\partial \bm p)
+ {\cal C}[f]/\gamma =0,
\end{equation}
where
\begin{equation}
  \label{eq:v-gamma}
  \bm v \equiv \bm p/(\gamma M)
\quad\mbox{and}\quad
\gamma\equiv(1-\bm v^2)^{-1/2}
\end{equation}
is the particle velocity and relativistic gamma-factor respectively.

The collision integral ${\cal C}[f]$ in Eq.~(\ref{eq:boltzmann-part}) gives the
collision frequency for all particles with momentum~$\bm p$ (near
space-time point $x$) in the rest frame of those particles, while
$C/\gamma$ is that frequency in the lab frame. Using equations of motion
(\ref{eq:particle-eom}) one can show that the Boltzmann equation
(\ref{eq:boltzmann-part}) (or (\ref{eq:boltz-noncov})), implies
continuity equation for the particle number current
\begin{equation}
  \label{eq:current}
  \partial_\mu j^\mu+  \int_{\bm p} {\cal C}[f]/\gamma = 0\,,
\quad\textrm{where}\quad
j^\mu \equiv  \int_{\bm p} f p^\mu/(M\gamma),
\end{equation}
with
\begin{equation}
  \label{eq:int-p}
  \int_{\bm p}\equiv\int \frac{d^3\bm p}{(2\pi)^3}\,.
\end{equation}
I.e., particle number can only be changed (if at all) by collisions.

The equation of motion for the scalar field $\sigma$ in the presence
of particles with distribution $f(x,p)$, following
Eq.~(\ref{eq:sigma-eom}), is given by
\begin{equation}
  \label{eq:sigma-f}
    \partial^2\sigma + dU/d\sigma + (dM/d\sigma) \int_{\bm p} f/\gamma = 0.
\end{equation}
Coupled equations~(\ref{eq:boltzmann-part}) and~(\ref{eq:sigma-f})
describe evolution of the particle distribution $f$ and the scalar
field $\sigma$. These equations are conceptually similar to 
Vlasov equations in electrodynamics. The difference is that the
classical field $\sigma$ is a Lorentz scalar. There is also certain
limited similarity with the nuclear mean-field
approach~\cite{PhysRevLett.54.289}.

Our goal is to extend the above formalism to the description of
fluctuations in the system. In application to linearized Boltzmann
equation this has been done by Fox and Uhlenbeck and
others~\cite{Fox1,Fox2,PhysRev.187.267,LoganKac}. Here we shall extend this
formalism to linearized Boltzmann-Vlasov type coupled
equations~(\ref{eq:boltzmann-part}) and~(\ref{eq:sigma-f}).

\subsection{Linearized equations}
\label{sec:linear}

For a given constant field $\sigma$, the Boltzmann equation has a
stationary solution, which is also constant in space, $f_\sigma(\bm
p)$, satisfying ${\cal C}[f_\sigma]=0$.
This is Boltzmann distribution for particles of mass $M(\sigma)$ at
arbitrary values of temperature $T$ and chemical potential $\mu$: 
\begin{equation}
f_\sigma(\bm p)=e^{\mu/T}\,e^{-\gamma(\bm p) M/T}\,.
\label{eq:f-sigma}
\end{equation}
The values of $T$ and
$\mu$ depend on the total particle number (if it is conserved by
collisions) and on total energy, if the system is closed, or by conditions of
equilibrium with the thermal bath if it is open.

The equilibrium value of $\sigma$ is determined by
\begin{equation}
  \label{eq:sigma-bar}
  dU/d\sigma + (dM/d\sigma)\int_{\bm p} f_\sigma/\gamma = 0.
\end{equation}
where the second term can be viewed diagrammatically 
as the contribution of a thermal tadpole.

We linearize the equations for $\sigma$ and $f$ by expanding around
their equilibrium value.  The deviation of $f$ from its equilibrium
value $f_\sigma$ will be parametrized, as usual, by function $h$:
\begin{equation}
  \label{eq:f-h}
  f= f_\sigma (1+h).
\end{equation}
The linearized Boltzmann equation then reads
\begin{equation}
  \label{eq:Boltzmann-linear}
  \dot h - \dot\sigma\,g/(\gamma T) +\bm{v\cdot\nabla} h + {\cal I}[h] = 0,
\end{equation}
where
\begin{equation}
  \label{eq:g-def}
  g\equiv dM/d\sigma
\end{equation}
and ${\cal I}[h]$ is the linearized collision
integral:
\begin{equation}
  \label{eq:collision-linearized}
  {\cal C}[f]=\gamma f_\sigma\,{\cal I}[h]+{\cal O}(h^2).
\end{equation}
Note, that both ${\cal C}$ and ${\cal I}$ depend on the local value of the field
$\sigma$ (through the dependence of the particle mass $M$), and we
used the property of the equilibrium distribution ${\cal C}[f_\sigma]=0$.

Shifting the notation for $\sigma$ so that $\sigma=0$ is the
equilibrium value (solution of Eq.~(\ref{eq:sigma-bar}))
 we can write the linearized equation (\ref{eq:sigma-f}) as
\begin{equation}
  \label{eq:sigma-linearized}
  \ddot\sigma -\bm\nabla^2\sigma + m^2\sigma + g \int_{\bm p} f_0\,h/\gamma=0\,.
\end{equation}
where we defined the ``in-medium'' mass $m$ of the the $\sigma$ field
quanta as
\begin{equation}
  \label{eq:m2-def}
  m^2 = m_0^2 + \frac{d}{d\sigma}
\left(g\int_{\bm p}f_\sigma/\gamma\right)_{\sigma=0}\,.
\end{equation}
with $m_0^2\equiv d^2U(0)/d\sigma^2$. The last term in
Eq.~(\ref{eq:m2-def}) can be recognized as the one-loop thermal
contribution to the vacuum mass $m_0$.

In the system we considered so far the dissipation (entropy increase) is
entirely due to the collision term ${\cal C}[f]$. In a more general, and more
realistic, case when the field $\sigma$ interacts with other particles
in a heat bath, one can describe the additional dissipation effects
adding a term $\Gamma_0\dot\sigma$ into the l.h.s. of Eq.~(\ref{eq:sigma-linearized}).

\subsection{Noise and its correlators}
\label{sec:noise-correlator}

Equations (\ref{eq:boltz-noncov}) and~(\ref{eq:sigma-f}) describe
evolution of functions $f$ and $\sigma$ averaged over the time
scale of many particle collisions. Fluctuations of $f$ and $\sigma$
can be characterized by correlation functions (also averaged over
 many collisions). 

In order to describe these fluctuations
 we follow the approach of Ref.~\cite{Fox2} and introduce
random noise terms. We shall determine the correlation functions of
these noise terms following Refs.~\cite{Fox1,Fox2,Lifshitz:v9,Lifshitz:v10}
in the linearized regime. For that purpose we shall cast equations
in the following first-order form:
\begin{subequations}\label{eq:linear-noise}
  \begin{align}
\label{eq:lin-noise-h}
    \dot h & - \pi\,g/(\gamma T) +\bm{v\cdot\nabla} h + {\cal I}[h] =
    \xi ;
\\
\label{eq:lin-noise-pi}
    \dot\pi & + \Gamma_0\pi -\bm\nabla^2\sigma + m^2\sigma + g
    \int_{\bm p}
    f_0\,h/\gamma=\eta ;
\\
\label{eq:sigma-pi}
    \dot\sigma & -\pi=0.
  \end{align}
\end{subequations}
where we introduced noises $\xi$ and $\eta$ and, for
generality, the additional dissipation term $\Gamma_0\pi$.
It is worth pointing out that the field $\sigma$ is stochastic with
our without the noise $\eta$ because of the couping to the particles
(last term on the l.h.s. of Eq.~(\ref{eq:lin-noise-pi})).


To determine the correlators of the noises $\xi$ and $\eta$, we use
fluctuation-dissipation relation. The probability distribution of the
fluctuating degrees of freedom~\cite{Einstein:1910,Lifshitz:v5}
\begin{equation}
  \label{eq:probability}
  {\cal P}[f,\sigma] \sim \exp S[f,\sigma]
\end{equation}
is determined by the entropy function
\begin{multline}
  \label{eq:entropy}
  S[f,\sigma]= \intx\left[
    -\int_{\bm p} f\left(\log f-1\right) 
\right.\\ \left.
- \frac1{T}\left(\frac{\pi^2}2 +
      \frac{(\bm\nabla\sigma)^2}2
      + U(\sigma) + \int_{\bm p}(M(\sigma)\gamma - \mu) f\right) 
  \right],
\end{multline}
where $T$ is the temperature of the external heat bath.  The first
term is the well-known Boltzmann entropy ($H$-function), while the
second term is simply $-(E-\mu N)/T$, where $E$ is the energy of the
system and $\mu$ is the chemical potential.  This term is the
contribution of external reservoir to the (fluctuations of)
entropy.

In the linear approximation, we can consider noise to be Gaussian, and
all non-trivial information about it to be in the
correlators such as $\langle\xi(x_1,p_1)\xi(x_2,p_2)\rangle$,
$\langle\eta(x_1)\eta(x_2)\rangle$ and
$\langle\xi(x_1,p)\eta(x_2)\rangle$. To determine these correlators we
expand the entropy to quadratic order (note cancellation of terms
linear in $h$ due to $\log f_\sigma = (\mu - M(\sigma)\gamma)/T$)
\begin{equation}
  \label{eq:S-quadratic}
  S^{(2)} = -\frac12\intx\left[
\int_{\bm p}f_0 h^2
 + \frac1{T}\left({\pi^2} +
   (\bm\nabla\sigma)^2
  +   m^2\sigma^2 \right) 
\right] ,
\end{equation}
where
\begin{equation}
  \label{eq:m-m_0}
   m^2 = m_0^2 - T\int_{\bm p}\frac{d^2 f_\sigma}{d\sigma^2}
\end{equation}
is the same ``in-medium'' mass of the $\sigma$ field quanta already
defined in Eq.~(\ref{eq:m2-def}), as can be verified by using
Eq.~(\ref{eq:f-sigma}).

We follow refs.~\cite{Fox1,Fox2,LoganKac} to define the
``entropy matrix'' (or, more precisely, operator) $\mathbb E$:
\begin{equation}
  \label{eq:E-def}
   \mathbb E
  \begin{pmatrix}
    h\\\pi\\\sigma
  \end{pmatrix}
= 
  \begin{pmatrix}
    f_0 h\\\pi/T\\\left(-{\bm\nabla}^2\sigma+m^2\sigma\right)/T
  \end{pmatrix}\,,
\end{equation}
so that
Eq.~(\ref{eq:S-quadratic}) can be written as
\begin{equation}
  \label{eq:ent-matrix}
   S^{(2)}  = - \frac12\bm h\bm\cdot \mathbb E \bm h,
\end{equation}
where $\bm h\equiv(h,\pi,\sigma)$ denotes the (infinitely dimensional) ``vector''
whose components are the degrees of freedom of the system, with the
scalar product defined as
\begin{equation}
(h,\pi,\sigma)\bm\cdot (h',\pi',\sigma')
\equiv
\intx
\left[ \int_{\bm p} hh' +  \pi\pi' + \sigma\sigma'\right].
\label{eq:ycdot}
\end{equation}

Similarly, equations~(\ref{eq:linear-noise}) can be also cast in matrix
(operator) form
\begin{equation}
  \label{eq:lin-noise-G}
  \dot {\bm h} + \mathbb G \bm h = \bm \xi,
\end{equation}
where
\begin{equation}
  \label{eq:G-def}
  \mathbb G
  \begin{pmatrix}
    h\\\pi\\\sigma
  \end{pmatrix}
=
  \begin{pmatrix}
 - \pi\,g/(\gamma T) +\bm{v\cdot\nabla} h + {\cal I}[h]
\\
  \Gamma_0\pi -\bm\nabla^2\sigma + m^2\sigma + g
    \int_{\bm p}
    f_0\,h/\gamma
\\
 -\pi
  \end{pmatrix}\,.
\end{equation}
Then the correlator of the noises, combined into a vector $\bm \xi = (\xi, \eta,
0)$,
can be expressed in terms of the matrix/operator $\mathbb Q$ defined as
\begin{equation}
  \label{eq:xixi-Q}
  \langle\bm\xi(t_1)\otimes \bm\xi(t_2)\rangle
= 2\mathbb Q\,\delta(t_1-t_2)\,,
\end{equation}
and given by the usual fluctuation-dissipation relation
 (see Refs.~\cite{Fox1,Fox2,Lifshitz:v9})
\begin{equation}
  \label{eq:QGE}
  2\mathbb Q = \mathbb G\mathbb E^{-1} + \mathbb E^{-1} \mathbb G^\dag.
\end{equation}
Using Eqs.~(\ref{eq:E-def}),~(\ref{eq:G-def}) and~(\ref{eq:QGE}) one
can now find $\mathbb Q$:
\begin{equation}
  \label{eq:Qh}
  2\mathbb Q
  \begin{pmatrix}
    h\\\pi\\\sigma
  \end{pmatrix}
=
\begin{pmatrix}
{\cal I}[h/f_0] + {\cal I}^\dag[h]/f_0\\2\Gamma_0 T\pi\\ 0
\end{pmatrix}
 =
 \begin{pmatrix}
   ({\cal K}+{\cal K}^\dag)[h]\\ 2\Gamma_0 T \pi\\ 0
 \end{pmatrix}
,
\end{equation}
where we defined operator ${\cal K}$ as
\begin{equation}
  \label{eq:K}
  {\cal I}[h] \equiv  {\cal K} [f_0 h]\,.
\end{equation}
One can show~\cite{Lifshitz:v10} that the operator ${\cal K}$ is self-adjoint for
elastic collisions, but we leave equations in a more general form.
Equation~(\ref{eq:Qh}) together with Eq.~(\ref{eq:xixi-Q}) 
translates into the following explicit expression for the
correlators:
\begin{subequations}\label{eq:noise-correlators}
  \begin{align}
    &\langle\xi(x_1, p_1)\xi(x_2, p_2)\rangle \nonumber\\
    &\qquad = ({\cal K}+{\cal K}^\dag)
    (2\pi)^3\delta^3( \bm p_1 - \bm p_2)\delta^4(x_1-x_2);
\label{eq:xi-xi}
\\
\label{eq:eta-eta}
&\langle\eta(x_1)\eta(x_2)\rangle =
    2\Gamma_0 T\delta^4(x_1-x_2);\\
\label{eq:xi-eta}
&\langle\xi(x_1, p_1)\eta(x_2)\rangle = 0.
  \end{align}
\end{subequations}
It is easy to recognize in eq.~(\ref{eq:xi-xi}) the 
generalization of the result of Ref.~\cite{Fox2}. One can also observe
that the interaction between the particles and the field $\sigma$ does
not manifest itself in any modification of the corresponding
noises. That should be expected given the physical origin of the
noise: collisions and the interaction
with the external reservoir. The correlations are local in coordinate
space, which also correctly reflects their origin.

\section{Stationary, equilibrium fluctuations}
\label{sec:equilibrium-fluc}


The quantity directly accessible by experimental measurement is a
two-particle correlator $\langle\delta f_{(1)}\delta f_{(2)}\rangle$.
Before we begin studying time evolution of fluctuations let us derive
the stationary, equilibrium value of the two-particle correlator and
compare with existing results.

To
linear order in fluctuations,
\begin{equation}
  \label{eq:f-sigma-h}
  f=f_\sigma(1+h) = f_0(1 + h - g\sigma/(\gamma T)) + {\cal O}(\sigma^2).
\end{equation}
The equal-time correlators of $h$ and $\sigma$ are contained in
the matrix elements of the correlator of $\bm h$ which {\em in
  equilibrium} are given by
\begin{equation}
  \label{eq:h-h}
  \langle \bm h \otimes \bm h\rangle = \mathbb E^{-1}
\end{equation}
according to (\ref{eq:ent-matrix}).
Using the explicit expression (\ref{eq:E-def})
for the components of the entropy matrix we thus find
\begin{multline}
    \label{eq:deltafx2}
    \langle\delta f_{(1)}\delta f_{(2)}\rangle = \langle f_{(1)} f_{(2)}\rangle -
     f_{0(1)}f_{0(2)}\\ 
     = f_{0(1)}f_{0(2)}
     \left\langle 
       \left(
         h- \frac{g\sigma}{\gamma T}
       \right)_{(1)}
       \left(
         h- \frac{g\sigma}{\gamma T}
       \right)_{(2)}
     \right\rangle \\
    =f_{0(1)}\delta_{(1,2)}+ \frac{g^2}{T}\frac{f_{0(1)}f_{0(2)}}{
      \gamma_{(1)} \gamma_{(2)}} D_{(1,2)}    \,,
\end{multline}
where subscripts $(1)$ and $(2)$ refer to the points in the phase
space $(\bm x_1, \bm p_1)$ and $(\bm x_2, \bm p_2)$ where the scripted
quantities are to be evaluated,
 $\delta_{(1,2)} = \delta^3(\bm x_1 -
\bm x_2)(2\pi)^3\delta^3(\bm p_1 -
\bm p_2)$ and $D_{(1,2)}=(-\bm\nabla^2+ m^2)^{-1}\delta^3(\bm
x_1-\bm x_2)$.
Integrating over $\bm x_{1,2}$ one obtains the known result for the
equilibrium fluctuations of particles coupled to classical scalar field
$\sigma$~ \cite{Stephanov:1999zu,Stephanov:2001zj}:
    \begin{equation}
    \label{eq:nu-correlator-equilibrium}
    V^{-1}\langle\delta\nu_{\bm p}\delta\nu_{\bm p'}\rangle =
    f_0\delta_{\bm p \bm p'} 
    +
    \frac{g^2}{m^2T}\frac{f_0}{\gamma}\frac{f_0'}{\gamma'} \,,
     \end{equation}
     where $V=\intx$ is the volume, 
$\delta_{\bm p \bm p'}\equiv(2\pi)^3\delta^3(\bm p -\bm p')$, and we
denoted the momentum space distribution as
\begin{equation}
  \label{eq:mom-density}
  \nu_{\bm p} = \intx\ f(\bm x,\bm p),
\quad\mbox{such that}\quad
 \int_{\bm p}
\nu_{\bm p} = N
\end{equation}
is the total number of the particles.

\section{Time evolution of fluctuations}
\label{sec:time-evol-fluc}

Now we want to consider the time evolution of fluctuations as the
parameters of the system, most importantly~$m$, change.

If we take the initial probability distribution for fluctuating
variables to be Gaussian, in a linear system the fluctuations will
remain Gaussian at all times. This can be verified directly, by
converting the generalized Langevin equations (\ref{eq:lin-noise-G})
into corresponding Fokker-Plank equation for the probability
distribution $P[\bm h;t]$:
\begin{equation}
  \label{eq:P-G}
  \dot P = \frac\partial{\partial \bm h}{\bm\cdot}
  \left(
    \mathbb G \bm h P + \mathbb Q \frac{\partial P}{\partial \bm h}
  \right).
\end{equation}
Parametrizing the probability using (time-dependent) operator $\bm\Sigma$:
\begin{equation}
  \label{eq:Sigma}
  P = (\det\bm\Sigma)^{-1/2}\exp\left[-\frac12\bm {h\cdot\Sigma^{-1} h}\right],
\end{equation}
and substituting into Eq.~(\ref{eq:P-G}), one finds equation for
$\bm\Sigma$:
\begin{equation}
  \label{eq:dotSigma}
  \dot{\bm\Sigma} = -\mathbb G \bm\Sigma - \bm\Sigma\mathbb G^\dag +
  2\mathbb Q  \,.
\end{equation}
Note that $\bm\Sigma=\mathbb E^{-1}$ is a stationary solution of this
equation, as expected from Eq.~(\ref{eq:h-h}) and the fact that
\begin{equation}\langle \bm h \otimes \bm h\rangle=\bm\Sigma.
\label{eq:hh-Sigma}
\end{equation}
This verifies the fluctuation-dissipation relation (\ref{eq:QGE}).

The equation~(\ref{eq:dotSigma}) for $(d/dt)\langle \bm h \otimes \bm h\rangle$ can
be also derived directly, by applying Eqs.~(\ref{eq:lin-noise-G})
and~(\ref{eq:xixi-Q}), without assuming Gaussianity (\ref{eq:Sigma}).

Equation~(\ref{eq:dotSigma}) can be formally integrated from initial time,
taken to be 0, to arbitrary time $t$:
\begin{equation}
  \label{eq:Sigma(t)}
  \bm\Sigma(t) =\mathbb V(t,0) \bm\Sigma(0) \mathbb V^\dag(t,0) 
+ 2 \int_0^t dt'\, \mathbb V(t,t')\mathbb Q \mathbb V^\dag(t,t')\,,
\end{equation}
where evolution operator $\mathbb V(t,t')$ satisfies
\begin{equation}
\dot {\mathbb V}
=-\mathbb G(t) \mathbb V,
\label{eq:dot-V}
\end{equation}
i.e.,
\begin{equation}
  \label{eq:V-G}
  \mathbb V(t,t') = {\cal T} \exp
  \left(
    -\int_{t'}^t dt''\,\mathbb G (t'')
  \right).
\end{equation}
Equation~(\ref{eq:dotSigma}) can be now used to study the time evolution
of the fluctuations, provided, of course, they remain small, so that
linear approximation is valid at all relevant times. 

A more useful equivalent form of equation (\ref{eq:dotSigma}) is
\begin{equation}
  \label{eq:dot-tilde-Sigma}
  \frac{d\tilde{\bm\Sigma}}{dt} = 
-\mathbb G\tilde{\bm\Sigma} - \tilde{\bm\Sigma}\mathbb G^\dag
- \frac{d(\mathbb E^{-1})}{dt}\,,
\quad\mbox{where}\quad
\tilde{\bm\Sigma}\equiv \bm\Sigma - \mathbb E^{-1}\,.
\end{equation}
In this form, and for $\mathbb E={\rm const}$, it describes relaxation
of $\bm\Sigma$ to its equilibrium value $\mathbb E^{-1}$. The
solution is given by
\begin{equation}
  \label{eq:tildeSigma(t)}
  \tilde{\bm\Sigma}(t) =\mathbb V(t,0) \tilde{\bm\Sigma}(0) \mathbb V^\dag(t,0) 
- \int_0^t \!dt'\,\mathbb V(t,t')\frac{d(\mathbb E^{-1})}{dt'} \mathbb
V^\dag(t,t')
,
\end{equation}
which is equivalent to Eq.~(\ref{eq:Sigma(t)}).

Determining the time-dependence more explicitly in the general case is
a complicated task, and should perhaps be part of numerical modeling
of a more realistic system. As an illustration of the use of
Eq.~(\ref{eq:tildeSigma(t)}) we shall address the important question of
the fate of the critical point fluctuations after chemical freezeout.
To prepare for this, we shall briefly discuss the way conservation of
particle number is reflected in the equations
(Section~\ref{sec:cons-part-numb}), and then
analyze the evolution of fluctuations for 
time-independent $\mathbb G$ and $\mathbb E$
(Sections~\ref{sec:const-coeff} and~\ref{sec:slowest-mode}).


\section{Conservation of particle number and fluctuations}
\label{sec:cons-part-numb}

Chemical freezeout is a moment in the history of a
heavy-ion collision fireball, when inelastic reactions become too
infrequent to modify the chemical composition of the system. In other words,
the number of particles of a given species is conserved during
subsequent evolution.

To model the evolution past the chemical freezeout, we shall require that
the collision integral ${\cal C}[f]$ conserves the particle
number. According to Eq.~(\ref{eq:current}) this requires
\begin{equation}\label{eq:C=0}
  \int_{\bm p} {\cal C}[f]/\gamma=0
\end{equation}
to be valid for all $f$. The linearized collision operator ${\cal I}$ in
Eq.~(\ref{eq:collision-linearized}) therefore obeys, for all $h$,
\begin{equation}
  \label{eq:MI=0}
  {\cal M}[{\cal I}[h]]=0
\end{equation}
For future convenience, we have introduced notation
\begin{equation}\label{eq:M-def}
{\cal M}[h]\equiv\frac{\int_{\bm p}f_0 h}{\int_{\bm p}f_0}
\end{equation}
for the average over equilibrium distribution $f_0$.
Imposing the condition that operator ${\cal K}$ defined in
Eq.~(\ref{eq:K}) is self-adjoint~\cite{Lifshitz:v10}, one can see that
Eq.~(\ref{eq:MI=0}) implies
\begin{equation}
  \label{eq:I-const}
  {\cal I}[{\rm const}]=0.
\end{equation}
In other words, operator ${\cal I}$ has a zero mode. This is also
evident from the fact that constant $h$ corresponds to changing the
value of $\mu$ in the equilibrium distribution~(\ref{eq:f-sigma}), and
that ${\cal C}[f_\sigma]=0$ for arbitrary $\mu$.


\section{Solving evolution equations with constant coefficients}
\label{sec:const-coeff}

The evolution operator $\mathbb V$ in Eq.~(\ref{eq:V-G}) for the
system with time-independent $\mathbb G$ can be written in the form:
\begin{equation}
  \label{eq:V-sum-lambda}
  \mathbb V(t,t') = \sum_\lambda e^{-\lambda (t-t')}\bm h_\lambda\otimes \hdual_\lambda\,,
\end{equation}
where the sum goes over all solutions of the following eigenvalue
system
\begin{equation}
  \label{eq:lambda-h-G-h}
  \lambda \bm h_\lambda = \mathbb G \bm h_\lambda\,,
\end{equation}
and vectors $\hdual_\lambda$ form
the dual (adjoint) basis with respect to the one formed by vectors  ${\bm h}_\lambda$, i.e., 
\begin{equation}
  \label{eq:bar-h-lambda}
  \hdual_{\lambda_i} \bm\cdot \bm h_{\lambda_j} = \delta_{ij}.
\end{equation}
If the system is also spatially homogeneous, it is convenient
to apply Fourier transformation with respect to the space
coordinate $\bm x$ to equations (\ref{eq:lambda-h-G-h}). Given the
definition of operator $\mathbb G$ in Eq.~(\ref{eq:G-def}), we
find:
\begin{subequations}\label{eq:lambda-eigensystem}
  \begin{align}\label{eq:lambda-h-q}
    -\lambda h & - \pi\,g/(\gamma T) + i\bm{v\cdot q} h + {\cal I}[h]
    = 0; \\\label{eq:lambda-pi-q} 
    -\lambda \pi & + \Gamma_0\pi + \bm
    q^2\sigma + m^2\sigma + g n_0 {\cal M}[h/\gamma]=0;\\
    -\lambda \sigma & -\pi=0;
  \end{align}
\end{subequations}
where we defined
\begin{equation}
  \label{eq:n0}
  n_0 \equiv \int_{\bm p} f_0
\end{equation}
-- the  equilibrium density of the particles.

\section{The slowest mode}
\label{sec:slowest-mode}

The slowest mode corresponding to the lowest eigenvalue of the
eigensystem (\ref{eq:lambda-eigensystem}) is of primary interest to us.
In this section we shall determine it.

\subsection{Zero mode}
\label{sec:zero-mode}

We begin by considering the simpler case $\bm q = 0$.
Since operator ${\cal I}$ has
a zero mode $h={\rm const}$ (\ref{eq:I-const}), let 
us separate it by writing
\begin{equation}
  \label{eq:h-tilde-h-0}
  h = \tilde h + \hmean,\quad\mbox{where}\quad \hmean\equiv{\cal M}[h].
\end{equation}

The zero eigenvalue $\lambda_0=0$ of Eqs.~(\ref{eq:lambda-eigensystem})
corresponds to the solution such
that $\tilde h=\pi=0$ (so that Eq.~(\ref{eq:lambda-h-q}) is trivial)
and
\begin{equation}
  \label{eq:h-lambda0-m2-sigma}
{\bm h}_{\lambda_0}:\quad  gn_0{\cal M}[1/\gamma]\hmean= -m^2\sigma
\qquad (\lambda_0=0).
\end{equation}
Note that this mode is predominantly $\sigma$ near the critical point
(i.e, $\bar h\to0$ as $m\to0$). This may appear surprising, since the
mode $\lambda_0$ is due to the particle number conservation, while
$\sigma$ is not a density of a conserved quantity. However, near the
critical point the fluctuations of particle number density are
dominated by their mixing with
$\sigma$~\cite{Hatta:2003wn,Stephanov:2004wx}, whose fluctuations
diverge. Eq.~(\ref{eq:h-lambda0-m2-sigma}) also shows that in the limit $g\to0$
at fixed $m$, the $\lambda_0$ mode is predominantly~$\bar h$, as it
should be if $\sigma$ is decoupled.

In order to find the dual vector $\hdual_{\lambda_0}$ we need some information
about all other modes $\bm h_\lambda$, since $\hdual_{\lambda_0}$ must
be orthogonal to them (\ref{eq:bar-h-lambda}). Applying operator
${\cal M}$ to equation~(\ref{eq:lambda-h-q}) and using
(\ref{eq:MI=0}), we find, at $\bm q=0$:
\begin{equation}
  \label{eq:hlambda-hmean-sigma}
{\bm h}_\lambda:\quad  \hmean = \frac gT {\cal M}[1/\gamma]\sigma
\qquad (\lambda\neq0).
\end{equation}
This equation contains all the information about the non-zero modes that
we need to determine $\hdual_{\lambda_0}$. 

In order to simplify subsequent linear algebra manipulations, we shall
define two convenient basis vectors:
\begin{align}
\label{h-hat}
  \hhat:&\quad
\tilde h=\pi=\sigma=0
\quad \mbox{and} \quad
\hmean = g{\cal M}[1/\gamma]/T;\\
\label{s-hat}
\hat{\bm\sigma}:&\quad
\tilde h=\pi=\hmean=0
\quad \mbox{and} \quad
\sigma=1.
\end{align}
In terms of these vectors, Eq.~(\ref{eq:hlambda-hmean-sigma}) means that, for any nonzero
eigenvalue $\lambda\neq0$, eigenmodes are given by
\begin{equation}
  \label{eq:h-lambda_i}
  \bm h_{\lambda} =  \hat{\bm \sigma} + \hhat
+ \mbox{(terms with $\hmean=\sigma=0$)}\,,
\end{equation}
while Eq.~(\ref{eq:h-lambda0-m2-sigma}) for  the zero mode can be written as
\begin{equation}
  \label{eq:h-lambda0-shat-hhat}
  \bm h_{\lambda_0} =  \deli^2 \hat{\bm \sigma} - {m^2}\hhat \,,
\end{equation}
where we defined
\begin{equation}\label{eq:deli-def}
 \deli^2 \equiv \frac{g^2n_0}{T}{\cal M}[1/\gamma]^2.
\end{equation}
Equations~(\ref{eq:h-lambda_i}) and~(\ref{eq:h-lambda0-shat-hhat}) 
together determine the orientation and the length of the dual vector
\begin{equation}
  \label{eq:hdual-lambda0}
  \hdual_{\lambda_0} = \frac1V\,\frac{ \hat{\bm \sigma} - \hdualhat }
{\deli^2+m^2}  \,,
\end{equation}
which satisfies the defining orthonormality conditions~(\ref{eq:bar-h-lambda}).
We defined another convenient vector, related to (\ref{h-hat}):
\begin{equation}
  \label{eq:h-dual-hat}
  \hdualhat = \frac{T}{\deli^2}f_0\hhat\,,
\end{equation}
such that $\hdualhat\bm\cdot\hhat=1$. One can also check that
 $\bm\theta_{\lambda_0}$ is the $\lambda=0$ eigenvector of
$\lambda\bm\theta_\lambda=\mathbb G^\dag\bm\theta_\lambda$, as it should be.

\subsection{Hydrodynamic mode and diffusion coefficient}
\label{sec:hydro-mode}

The mode $\lambda_0$, corresponding to conservation of the
particle number, is hydrodynamic in the sense that, for small~$\bm
q$, $\lambda_0={\cal O}(\bm q^2)$.
 The ratio $\lambda_0/\bm q^2=D$ defines the
corresponding diffusion coefficient $D$, which we can extract from
equations~(\ref{eq:lambda-eigensystem}). 

As we did deriving Eq.~(\ref{eq:hlambda-hmean-sigma}) for nonzero modes at $\bm
q=0$, let us apply operator ${\cal M}$ to
Eq.~(\ref{eq:lambda-h-q}). Now, at $\bm q\neq 0$, we find
\begin{equation}
  \label{eq:h-sigma-mean}
  \lambda\hmean 
- i {\cal M}[\bm{v\cdot q}\tilde h]
=\lambda \frac{g}{T}{\cal M}[1/\gamma]\sigma\,.
\end{equation}
We need now to express $\tilde h$ in terms of $\hmean$ and $\sigma$
using equation~(\ref{eq:lambda-h-q}).
Since $\tilde h\to0$ as $\bm q\to0$, one can see that $\tilde h$
must begin at order $\bm q$. Keeping in Eq.~(\ref{eq:lambda-h-q}) only
terms of ${\cal O}({\bm q})$ we obtain:
\begin{equation}
  \label{eq:Ih-h}
  i\bm{ v\cdot q}\,\hmean + {\cal I}[\tilde h] = 0\,. 
\end{equation}
We would need to invert operator ${\cal I}$ to express $\tilde h$ in
terms of $\hmean$. For generic operator ${\cal I}$ we shall define
function $\psi(\bm v^2)$, which solves the equation
\begin{equation}
  \label{eq:psi-def}
  {\cal I}[\bm v\, \psi] = \bm v\,.
\end{equation}
The fact that solution can be found in this form follows from isotropy
of the collision operator and equilibrium distribution function.
In terms of $\psi$, the solution to equation~(\ref{eq:Ih-h}) is given
by
\begin{equation}
  \label{eq:h-hmean-psi}
  \tilde h = -i\bm {q\cdot v}\,\psi \hmean \,.
\end{equation}
Substituting this into Eq.~(\ref{eq:h-sigma-mean}) we find, 
instead of Eq.~(\ref{eq:hlambda-hmean-sigma}),
\begin{equation}
  \label{eq:h-lambda0-D-q2}
  \hmean (\lambda_0 - D_0 \bm q^2) = \lambda_0\frac {g}{T}{\cal M}[1/\gamma] \sigma\,, 
\end{equation}
where we denoted by $D_0$
\begin{equation}
  \label{eq:D_0}
  D_0 \equiv \frac13 {\cal M}[\bm v^2\psi]
\end{equation}
the diffusion coefficient for the particle gas with {\em fixed} mass
(the limit $g\to0$). Putting together Eq.~(\ref{eq:h-lambda0-D-q2}) 
and~(\ref{eq:h-lambda0-m2-sigma}), which remains valid to the order in $\bm q^2$ we
need, we find
\begin{equation}
  \label{eq:D-D_0}
  \lambda_0=D\bm q^2 + {\cal O}(\bm q^4),
\quad\mbox{where}\quad
  D = \frac{m^2}{\deli^2+m^2} D_0.
\end{equation}
The fact that
$D\to0$ as $m^2\to0$ is to be expected on general grounds from the
hydrodynamic relation $D=\bar\sigma/\chi$~\cite{Hohenberg:1977ym}, where $\bar\sigma$ is the
conductivity, and $\chi$ is the susceptibility of the particle number,
and the fact that $\chi\sim
1/m^2$~\cite{Hohenberg:1977ym,Son:2004iv}. 
Within our microscopic approach:
\begin{equation}
  \label{eq:chi}
  \chi=\frac{\langle\delta N^2\rangle}{VT}
=\frac{n_0}{T}\,\frac{\deli^2+m^2}{m^2}
\end{equation}
according to Eq.~(\ref{eq:deltaN2/N}) which we encounter later.

\section{Faster modes}
\label{sec:non-zero-modes}


This section is a slight detour from the main thread of the paper. We
have already accumulated all information about the zero and even non-zero
modes, Eqs.~(\ref{eq:h-lambda0-m2-sigma}),~(\ref{eq:hlambda-hmean-sigma}), that we need to
study the ``memory'' effect in fluctuations (Section
\ref{sec:chem-to-kin}). However, it might still be interesting to look at the
structure of the {\em non-zero} modes in more detail, to understand better
the properties of the system of equations we are solving.

So far we have not used any information about the linear collision
operator $\cal I$ beyond the conservation of the particle number and
isotropy. 
For the sake of analytic transparency, and within this Section only,
we shall assume here that all eigenvalues, but one, of the operator
${\cal I}$ are equal to the same value $\tau^{-1}$, which has the
meaning of an average relaxation rate. The exception is the zero
eigenvalue, corresponding to the condition (\ref{eq:MI=0}). This
approximation is well known and is due to
Refs.~\cite{Anderson1974466,BGK.PhysRev.94.511} (see also
Ref.~\cite{CercignaniKremer2002}).

Operator ${\cal I}$ should also respect the condition that  operator ${\cal K}$,
defined by (\ref{eq:K}) is self-adjoint. All the above conditions are
satisfied by the operator
\begin{equation}
  \label{eq:I-tau}
  {\cal I}[h]=\tau^{-1}
  \left(
    h - {\cal M}[h]
  \right)\,.
\end{equation}

Another simplification we shall adopt in this section only corresponds to
assuming $\tau^{-1}\gg\Gamma_0,m$, i.e., that $\tau^{-1}$ is much faster
than any other rate in the problem.

We emphasize that, although these approximations are physically
sensible, they are only used here to make a transparent 
analytic treatment possible, illustrating the properties of the system
we study.


Substituting Eq.~(\ref{eq:hlambda-hmean-sigma}) back into
(\ref{eq:lambda-h-q}) and solving for~$\tilde h$ we find, at $\bm q\to0$,
\begin{equation}
  \label{eq:h-tilde-sigma}
  \tilde h = \frac{\lambda\tau}{1-\lambda\tau}\frac{g}{T}
\left({\cal M}[1/\gamma]-1/\gamma\right)\sigma\,.
\end{equation}
Now substituting $h$ given by
Eqs.~(\ref{eq:h-tilde-h-0}),~(\ref{eq:hlambda-hmean-sigma})
and~(\ref{eq:h-tilde-sigma}) into Eq.~(\ref{eq:lambda-pi-q}), we find
the equation determining the eigenvalues $\lambda$:
\begin{equation}
  \label{eq:lambda-ev-eq}
  \lambda^2-\lambda \Gamma_0 + m^2 
+ \frac{\deli^2-\lambda\tau\delii^2}{1-\lambda\tau}=0.
\end{equation}
where we used (\ref{eq:deli-def}) and defined also
\begin{equation}
  \label{eq:delii-def}
  \delii^2 \equiv \frac{g^2n_0}{T}{\cal M}[1/\gamma^2].
\end{equation}

Equation~(\ref{eq:lambda-ev-eq}) has three roots. For the scale hierarchy
we consider, $\tau^{-1}\gg\Gamma_0,m$, there are two roots of order
$\Gamma_0$ or $m$ and one root of order $\tau^{-1}$. The two smaller roots, to
leading order in $\tau$, satisfy the quadratic equation
\begin{equation}
  \label{eq:lambda-2-quadratic}
  \lambda^2-\lambda \Gamma + \tilde m^2=0\,,
\end{equation}
where
\begin{align}
      \label{eq:m-tilde-def}
      &\tilde m^2 \equiv m^2 + {\deli}^2;\\
      \label{eq:Gamma-tilde-def}
      &\Gamma \equiv \Gamma_0 + \delii^2\,\tau;
\end{align}
and thus
\begin{equation}
  \label{eq:lambda-1-2}
  \lambda_{1,2}=-\Gamma/2\pm i\sqrt{\tilde m^2-(\Gamma/2)^2}.
\end{equation}

At this point one can see that $\tilde m$ is the rest mass (pole
mass) of the quasiparticle $\sigma$. It is different from the (static)
screening mass $m$, and does not vanish at the 
critical point~\cite{Scavenius:2000qd,Fujii:2003bz,Son:2004iv}, where
 $m\to0$. 

Also, Eq.~(\ref{eq:Gamma-tilde-def}) shows that 
the full dissipation rate $\Gamma$ contains
contribution $\delii^2\tau$ from the interaction of $\sigma$ with the
particles. In principle, one could start with $\Gamma_0=0$ (closed
system) and
consider the particle collisions to be the only source of the
dissipation.

The third eigenvalue is given, to the leading nontrivial order in
$\tau$, by
\begin{equation}
  \label{eq:lambda-3}
  \lambda_3 = \tau^{-1} + (\deli^2 - \delii^2)\tau.
\end{equation}
(Applying Cauchy-Bunyakovsky-Schwarz inequality to
Eqs.~(\ref{eq:deli-def}) and~(\ref{eq:delii-def}), one can see that
$\deli<\delii$.)

Finally, any function
$h_\perp$ satisfying
\begin{equation}
  \label{eq:h_perp}
  {\cal M}[h_\perp]=0
\quad\mbox{and}\quad
{\cal M}[h_\perp/\gamma]=0
\end{equation}
solves the eigensystem with $\sigma=\pi=0$ and
$\lambda=\tau^{-1}$. The linear (eigen)space defined by Eqs.~(\ref{eq:h_perp})
is infinitely dimensional and, correspondingly, $\lambda=\tau^{-1}$ is an
infinitely degenerate eigenvalue. This degeneracy is not lifted because
modes $h_\perp$ do not mix with the modes corresponding to eigenvalues
$\lambda_i$, $i=0,1,2,3$. This is a convenient feature of the
Anderson-Witting approximation (\ref{eq:I-tau}).

Returning to the $\lambda_0$ mode, with operator ${\cal I}$ in the form given by
Eq.~(\ref{eq:I-tau}), equation (\ref{eq:psi-def}) can be solved:
$\psi=\tau$, and $D_0={\cal M}[\bm v^2]\tau/3$.

\section{From chemical to kinetic freezeout}
\label{sec:chem-to-kin}

\subsection{Preliminaries}
\label{sec:prelim}

As an application of the formalism, let us consider the following
long-standing problem. Assuming that the chemical freezeout occurred
near the critical point, how much of the fluctuation signal survives
until kinetic freezeout? A more precise and detailed answer to this
problem will likely require a numerical simulation. Here we want to
illustrate the mechanism, and make a simple estimate of the effect. To
that end we shall make several simplifying assumptions, in order to
maintain analytical control. In essence, we
shall assume that separation of different relaxation time scales is
sufficiently large for us to be able to focus on only the most
relevant modes.

Chemical freezeout is characterized by ``freezing'' of {\em inelastic}
reactions. This means that the number of each individual particles is
conserved (particles in the same isospin multiplet could be considered
as different internal states of the same particle, to allow for
quasielastic collisions). This, in turn, means that any measure of
fluctuations of a conserved number of particles should not
change. More precisely, it can only change by diffusion, which we
shall assume here to be the slowest scale in the problem (i.e., we
work in the $\bm q\to0$ limit).

On the other hand, fluctuations such
as those of mean~$p_T$, which is not a conserved
quantity, must evolve between chemical and kinetic
freezeout, at which point they are ``frozen'' and eventually observable. 
The form and the amount of this evolution we shall now discuss. 

During the interval between chemical and kinetic freezeout the typical
time scale,  $\tau_e$, of the evolution of the
system is much slower than the inverse elastic collision rate
$\tau$ and the scales $\Gamma^{-1}$ associated
with the relaxation of the $\sigma$ field. 
In a realistic heavy-ion
collision $\tau_e={\cal O}(10-20)$ fm (order of fireball size), while
$\Gamma^{-1},\,\tau= {\cal O}(0.5-2)$ fm (typical hadronic
scales). Thus we shall assume $\tau_e\gg\tau,\Gamma^{-1}$.

In order to be able to
obtain analytically tangible solution we shall take into account the
effect of the change of only one parameter: $m$ -- the screening mass
of $\sigma$. Since the fluctuations are singular as $1/m^2$ near the
critical point, the effect of change of $m$ could be assumed to be
dominant, compared with the change of, e.g., equilibrium distribution
functions $f_0$ (e.g., via change of $T$), which we shall consider
fixed, for simplicity. As a concrete example, one could consider
evolution of $m$ determined by the model in Ref.~\cite{Berdnikov:1999ph}. As
we shall see, the actual time dependence of $m$ will not matter,
as long as it is slow, which is helped by critical slowing
down~\cite{Berdnikov:1999ph}.

The physically reasonable assumptions spelled out above are
needed to make the analytic results attainable and usefully
transparent. These assumptions can be relaxed, e.g., via a
numerical simulation, at the expense of analytic control. Our main
purpose here is to illustrate the mechanism in the most transparent
way possible.

\subsection{Evolution of fluctuations}
\label{sec:evolution-fluctuations}

We begin by determining the evolution operator $\mathbb V(t,t')$. If
we choose the interval, $t-t'$, so small that we could neglect the change of
$\mathbb G$ (i.e., change of $m$) and consider it constant, then we could integrate
Eq.~(\ref{eq:V-G}) and obtain $\mathbb V$ given by Eq.~(\ref{eq:V-sum-lambda}).

If the interval $t-t'$ is also large compared to relaxation
scales $\lambda_i^{-1}$, for all $i\neq 0$, only the term
corresponding to the zero mode $\lambda=\lambda_0={\cal O}(\bm q^2)$
will survive in Eq.~(\ref{eq:V-sum-lambda}):
\begin{equation}
  \label{eq:V-lambda_0}
  \mathbb V(t,t') = e^{-\lambda_0(t-t')}\bm h_{\lambda_0}\otimes
  \hdual_{\lambda_0} + \mbox{(exp. small terms)}\,.
\end{equation}

In order to extend this result to longer time intervals over which the
change of $\mathbb G$ cannot be neglected, we use the property
\begin{equation}
  \label{eq:VVV}
  \mathbb V(t,t') 
  =  \mathbb V(t,t_n)\ldots \mathbb V(t_2,t_1) \mathbb V(t_1,t')
\end{equation}
and subdivide $t-t'$ into smaller intervals satisfying $\lambda_i^{-1}\ll
t_n-t_{n-1}\ll\tau_e$. 
Our assumption of scale hierarchy is needed
to make such a choice possible. Using Eq.~(\ref{eq:V-lambda_0}) we then
find
\begin{multline}
  \label{eq:V-hhh}
  \mathbb V(t,t') = e^{-\int_{t'}^t\lambda_0 dt}\,
\bm h_{\lambda_0}(t)\otimes\hdual_{\lambda_0}(t')\,
\\
\times(\hdual_{\lambda_0}(t)\bm\cdot\bm h_{\lambda_0}(t_n))
\ldots
(\hdual_{\lambda_0}(t_1)\bm\cdot\bm h_{\lambda_0}(t'))\,.
\end{multline}
In order to evaluate dot products in Eq.~(\ref{eq:V-hhh}), we use
explicit form of eigenvectors $\bm h_{\lambda_0}$ and
$\hdual_{\lambda_0}$, given by
Eqs.~(\ref{eq:h-lambda0-shat-hhat}),~(\ref{eq:hdual-lambda0}), and find, e.g.,
\begin{equation}
  \label{eq:hbar1-dot-h2}
  \hdual_{\lambda_0}(t_1)\bm\cdot\bm h_{\lambda_0}(t') = 
\frac{\deli^2+m^2(t')}{\deli^2+m^2(t_1)}\,.
\end{equation}
There is a string of such factors in Eq.~(\ref{eq:V-hhh}) and,
multiplying them successively, one finds that all
but the first and the last factor $\deli^2+m^2(t)$ cancel, leaving
\begin{multline}
  \label{eq:V-hh}
   \mathbb V(t,t') = e^{-\int_{t'}^t\lambda_0 dt}\,
\bm h_{\lambda_0}(t)\otimes\hdual_{\lambda_0}(t')
\frac{\deli^2+m^2(t')}{\deli^2+m^2(t)}\\
= e^{-\int_{t'}^t\lambda_0 dt}\,
\bm h_{\lambda_0}(t)\otimes\hdual_{\lambda_0}(t)\,.
\end{multline}
The only dependence on the initial time $t'$ remains in the
exponentially decaying prefactor. Since $\lambda_0=D \bm q^2$, this
prefactor is close to unity near the limit we have been working in:
$\bm q \to 0$ (the fact that $D\sim m^2\to0$ near the critical point
also helps).  In general, the importance of the prefactor depends on
the size of the region over which the fluctuations are measured. Since
$1/\lambda_0$ is an estimate of the time, $\tau_D$, it takes for a fluctuation to
diffuse over this region, the factor can be estimated roughly
as $\exp[-(t-t')/\tau_D]$. Below we shall consider the case when the region is
large enough, so that $\tau_D\gg\tau_e$.

We are now ready to apply Eq.~(\ref{eq:tildeSigma(t)}). We shall take
the initial time $t=0$ to be the time of chemical freezeout, and the
final time $t=\tk$ the time of kinetic freezeout. At chemical
freezeout, $t=0$, the fluctuations are equilibrated and
$\tilde\Sigma=0$.  Thus at kinetic freezeout, $t=\tk$,
equation~(\ref{eq:tildeSigma(t)}) gives, upon integration,
\begin{multline}
  \label{eq:tilde-Sigma(t_k)}
  \tilde{\bm\Sigma}(\tk) 
=\bm h_{\lambda_0}(\tk)\otimes\hdual_{\lambda_0}(\tk)\,\\
\times \left(\mathbb E^{-1}(\tc)  - \mathbb E^{-1}(\tk)\right) \,
\bm \hdual_{\lambda_0}(\tk)\otimes \bm h_{\lambda_0}(\tk)\,.
\end{multline}
A shorter way to derive Eq.~(\ref{eq:tilde-Sigma(t_k)}) is to observe
that the actual time-dependence of $m(t)$ is not important, as long as
it is faster than the diffusion: $\tau_D\gg\tau_e$ (but $\tau_e\gg\lambda_i^{-1}$). Choosing $m(t)$ to
have an (almost) instantaneous step from $\mc$ to $\mk$, and constant at all
other times one can then find solution (\ref{eq:tilde-Sigma(t_k)})
using equation~(\ref{eq:tildeSigma(t)}) with initial condition
$\tilde\Sigma(0)=\mathbb E^{-1}(\tc) - \mathbb E^{-1}(\tk)$.

Taking into account Eq.~(\ref{eq:E-def}), 
which for convenience we write, using
notations $\mathds 1_h$ for the unit operator  $\mathds 1_h [h]=h$ and
$\hat{\bm \pi}$ for the basis vector $(h=0,\pi=1,\sigma=0)$,%
\begin{equation}\label{eq:E-1}
\mathbb E^{-1}(t) = f_0^{-1} \mathds 1_{h} 
+  T \hat{\bm\pi}\otimes\hat{\bm\pi}
+ \frac T{m^2(t)}\hat{\bm\sigma}\otimes\hat{\bm\sigma}\,,
\end{equation}
together with Eqs.~(\ref{eq:h-lambda0-shat-hhat}) and~(\ref{eq:hdual-lambda0}),
we find for $\bm\Sigma = \mathbb E^{-1} +
\tilde{\bm\Sigma} $ at kinetic freezeout time
\begin{multline}
    \label{eq:Sigma(t_k)}
    {\bm\Sigma}(\tk) = 
    \mathbb E^{-1}(\tk) 
    \\
    +
    \left(\frac T{\mc^2} - \frac T{\mk^2}\right)
    \frac{ \deli^2 \hat{\bm \sigma}-\mk^2\hhat }
    {\deli^2+\mk^2}
    \otimes
    \frac{ \deli^2 \hat{\bm \sigma}-\mk^2\hhat }
    {\deli^2+\mk^2}\,,
\end{multline}
where $\mck=m(\tck)$ is the value of the $\sigma$ screening mass at
chemical/kinetic freezeout. The last term, containing $1/\mc^2$, is the
``memory'' effect, due to the freezing out of conserved particle
number fluctuations.

\subsection{Two-particle correlator and ``memory''}
\label{sec:two-particle-correlators}


In order to translate Eq.~(\ref{eq:Sigma(t_k)}) into observed
fluctuations, we should recall that $\bm\Sigma=\langle\bm h\otimes\bm
h\rangle$ and apply Eqs.~(\ref{eq:deltafx2}) (all but the last equality) to
calculate the 2-particle correlator. The fluctuations of the momentum space
distribution of particles~(\ref{eq:mom-density}) at
kinetic freezeout are thus given by
  \begin{multline}
    \label{eq:nu-correlator}
    V^{-1}\langle\delta\nu_{\bm p}\delta\nu_{\bm p'}\rangle =
    f_0\delta_{\bm p \bm p'}
    +
    \frac{g^2}{\mk^2T}\frac{f_0}{\gamma}\frac{f_0'}{\gamma'} 
\\
+
    \frac{g^2}{T}\left(\frac1{\mc^2}-\frac1{\mk^2}\right) f_0f_0'
\\\times
    \frac{\deli^2/\gamma+\mk^2{\cal M}[1/\gamma]}{\deli^2+\mk^2} \cdot
    \frac{\deli^2/\gamma'+\mk^2{\cal M}[1/\gamma]}{\deli^2+\mk^2}\,,
  \end{multline}
  which should be compared to Eq.~(\ref{eq:nu-correlator-equilibrium})
  with $m=\mk$.  To check that the effect of the additional term is to
  preserve the particle number (multiplicity) fluctuation at the value
  it attained at {\em chemical} freezeout, let us calculate that
  fluctuation by integrating Eq.~(\ref{eq:nu-correlator}) over momenta
  $\bm p$ and $\bm p'$ and using
  Eq.~(\ref{eq:mom-density}). Normalizing by the total number $\langle
  N\rangle=n_0V$ for convenience, we find
\begin{multline}
     \label{eq:deltaN2/N}
    \frac{\langle(\delta N)^2\rangle}{\langle N\rangle}
   = 1 
+ \frac{\deli^2}{\mk^2}
+\left(\frac1{\mc^2}-\frac1{\mk^2}\right)\deli^2
    = 1+\frac{\deli^2}{\mc^2}\,,
\end{multline}
where we used definition (\ref{eq:deli-def}). We see that, as
expected, the effect of the ``memory'' term is to keep multiplicity
fluctuations from changing after chemical freezeout.

The effect which is less obvious is that the ``memory'' term also
contributes to fluctuations of quantities which are {\em not} conserved. We
shall keep discussion as general as possible, but to be less
abstract, we shall consider fluctuations of mean transverse momentum $p_T$
per  event, which is one of the most common ``intensive'' measures of
fluctuations. This fluctuation can be also expressed via the
correlator~(\ref{eq:nu-correlator}) (see, e.g., Ref.~\cite{Stephanov:1999zu}):
\begin{equation}
  \label{eq:pt2}
  \langle(\delta p_T)^2\rangle = \frac1{\langle N\rangle^{2}}
\int_{\bm p}\int_{\bm p'} 
(p_T - \bar p_T)(p_T' - \bar p_T)
\langle\delta\nu_{\bm p}\delta\nu_{\bm
  p'}\rangle\,,
\end{equation}
where we defined
\begin{equation}
  \bar p_T \equiv {\cal M}[p_T].
\label{eq:bar-pt-def}
\end{equation}
Normalizing by $\langle N\rangle$
to remove trivial system-size scaling, we find
  \begin{multline}
    \label{eq:delta-p2-memory}
    \langle N\rangle \langle(\delta p_T)^2\rangle = {\cal M}\left[(p_T-\bar
    p_T)^2\right]
\\
+
  \frac{g^2n_0}{T}{\cal M}\left[(p_T-\bar p_T)/\gamma\right]^2
\left(
\frac{1-r_m}{\mk^2} + \frac{r_m}{\mc^2}
\right)\,,
  \end{multline}
where we 
introduced
\begin{equation}
  \label{eq:corr-factor}
  r_m=\left(\frac{\deli^2}{\deli^2+\mk^2}\right)^2
\,.
\end{equation}
Eq.~(\ref{eq:delta-p2-memory}) shows that the critical 
contribution ${\cal O}(1/\mc^2)$ can, under certain conditions, survive through the hadronic
rescattering stage until kinetic freezeout. 
Compared to the value at chemical freezeout, the ${\cal O}(1/\mc^2)$ term
is attenuated by the factor~$r_m$ (\ref{eq:corr-factor}) which, if the
$\sigma$ screening mass at kinetic freezeout, $\mk$, is
of order $\deli$ or smaller, is a non-negligible fraction of unity.

\subsection{Estimating the "memory" factor}
\label{sec:estimates-factor}

Let us now estimate the ``memory''
factor~(\ref{eq:corr-factor}). The value of
$r_m$ depends quite strongly on the ratio of $\mk$ to $\deli$. 
For fluctuations to survive, $\mk/\deli$ cannot be large.

The estimate for $\deli$ can be made using Eq.~(\ref{eq:deli-def}). In
order to do this correctly we need to generalize our analysis to
include more than one species of particles: nucleons (2 spin and 2
isospin states), pions, etc. We
then find that the expression for $r_m$ in
equation~(\ref{eq:corr-factor}) still holds, with $\deli^2$ receiving
contributions from all species:
\begin{equation}
  \label{eq:m-tilde-full}
  \deli^2 = \deli_{\rm nucleons}^2 + \deli_{\rm pions}^2 + \ldots\,.
\end{equation}
Choosing, for example, top SPS energy freezeout conditions $T=168$ MeV
and $\mu_B=266$ MeV~\cite{BraunMunzinger:1995bp}, we find for
the contribution of nucleons
$\deli_{\rm nucleons}\approx 430.\,(g_{p}/10.)\mbox{ MeV}$. We take
$g_p\approx m_p/f_\pi\sim 10$ as an estimate of the coupling of
$\sigma$ to protons. 

The estimate for the contribution of pions is
$\deli_{\rm pions}\approx 110.\,(g_\pi/2.)\mbox{ MeV}$, where
$g_\pi\approx G/m_\pi\sim 2.$, using the estimate for $G$ from
Ref.~\cite{Stephanov:1999zu}. The estimates for the contribution of
antinucleons and kaons are similarly small, compared to $\Delta_{\rm
  nucleons}$. Summation in quadratures increases the estimate for
$\Delta$ by less than 10\% over $\Delta_{\rm nucleons}$:
$\deli\approx 460$ MeV.

Thus, 
at top SPS energy, the critical $p_T$ fluctuations
survive at least half as well as the particle multiplicity fluctuations
($r_m>1/2$) until kinetic freezeout, if the $\sigma$ screening mass at the freezeout
does not exceed $\mk<\sqrt{\sqrt2-1}\,\deli\approx 300$ MeV.

\section{Summary, discussion and outlook}

In summary, we introduced an approach to studying time-dependent
quasi-stationary fluctuations near QCD critical point by combining
{\em stochastic} Boltzmann equation with an equation of motion for a
scalar field, describing the ``soft'' critical mode. 
We obtained the general solution of the linearized system
and studied its relaxation modes. We focused on the slowest
(diffusion) mode and analyzed its effect on the evolution of
fluctuations after chemical freezeout. 

One of the consequences of our analysis is the following
prediction. Under the conditions that particle number fluctuations are
frozen after chemical freezeout, the fluctuations of {\em
  non-conserved} quantities, such as, e.g., mean $p_T$, are also
preserved over time scales longer than collisional relaxation time
$\tau$. The strength of this effect crucially depends on the ratio of
the $\sigma$ screening mass $\mk$ at kinetic freezeout to $\deli$ (see
Eq.~(\ref{eq:corr-factor}) and Section~\ref{sec:estimates-factor}).

In other words, while for the multiplicity fluctuations to be preserved
after chemical freezeout the kinematic window of acceptance must be
large enough~\cite{Shuryak:2000pd,Koch:2008ia}, for the $p_T$
fluctuations to be preserved, additional condition is necessary:
$m_k<\Delta$.  We find that, e.g., at top SPS energies, $p_T$
fluctuations can survive the hadronic rescattering at least half as
well as the particle multiplicity fluctuations for $\mk< 300$ MeV.

The origin of this effect is the mixing between the critical mode
$\sigma$ and the conserved particle number density (see discussion
after Eq.~(\ref{eq:h-lambda0-m2-sigma})). E.g., when $\deli\gg m$, the
mode ${\bm h}_{\lambda_0}$, which is kept from relaxing by the
particle number conservation, is almost the same as~$\sigma$,
Eq.~(\ref{eq:h-lambda0-shat-hhat}). The
fluctuations of~$\sigma$ involved in $\bm h_{\lambda_0}$ must keep the
magnitude they reached at chemical freezeout, contributing the term $\sim
1/\mc^2$ into Eqs.~(\ref{eq:deltaN2/N})
and~(\ref{eq:delta-p2-memory}).  While multiplicity fluctuations in
Eqs.~(\ref{eq:deltaN2/N}) are frozen, the $p_T$ fluctuations evolve,
with contribution of the mode $\bm h_{\lambda_0}$ decreasing with
increasing $m$ as the factor $r_m$.

At the same time, the fluctuations of $\sigma$ {\em alone}, with
particle number fixed (i.e., obeying
Eq.~(\ref{eq:hlambda-hmean-sigma})), equilibrate on a short time
scale, $\Gamma^{-1}$, tracking the evolution of $m$. This
equilibrated mode of fluctuations contributes $1/\mk^2$ term into
Eq.~(\ref{eq:delta-p2-memory}).

In this paper we focused on fluctuations of one particle species,
treating the rest of the hadron gas as a heat bath. This
simplification allowed us to follow the evolution of fluctuations
analytically and expose the mechanism behind the ``memory'' effect in
the most transparent way.  This analysis could be generalized to the
case of multiple particle species, carrying (different values of) the
same conserved charge, as well as the case of multiple conserved
quantities (baryon number, isospin, etc.). Taking into account
fluctuations of conserved energy and momentum would be necessary, for
example, to obtain correct $m\to0$ scaling of the diffusion
coefficient \cite{Son:2004iv,Hohenberg:1977ym}. We leave this to
future work.

We also neglected the effects of quantum statistics for 
simplicity. Although these are relatively
small under realistic conditions (few percent, as estimated by mean
occupation numbers ${\cal M}[f_0]$), this approximation could be
removed. For the most part this would require replacing the equilibrium
distribution in Eq.~(\ref{eq:f-sigma}) with Bose-Einstein or
Fermi-Dirac distribution and factors $f_0$ in equations such as
Eq.~(\ref{eq:nu-correlator}) with $f_0(1\pm f_0)$. This would also
imply that the collision integral ${\cal C}[f]$ has Uehling-Uhlenbeck
form~\cite{PhysRev.43.552}. The influence functional 
method~\cite{Feynman:1963fq,Greiner:1996dx} could be used to derive the
corresponding equations.

We would like to stress that, although we did use relaxation-time
approximation to obtain more explicit formulas for {\em non-zero}
modes $\lambda$ in Section~\ref{sec:non-zero-modes}, the results
pertaining to the ``memory'' effect, which rely on the properties of
the zero mode $\lambda_0$ studied in Section~\ref{sec:slowest-mode}, are
valid beyond relaxation-time approximation.


A numerical simulation of the stochastic
equations~(\ref{eq:linear-noise}) should allow to take into account
more detailed properties of the heavy-ion collision evolution, such as
inhomogeneity, anisotropy and flow. The evolution of the $\sigma$ mass
$m$ can be described self-consistently, using equation
(\ref{eq:m-m_0}), conceptually reminiscent of nuclear mean-field
approach~\cite{PhysRevLett.54.289}, or disoriented chiral condensate
studies~\cite{Rischke:1998qy,Bettencourt:2001xd}.

We also deliberately limited our analysis to linearized regime and
focused on quadratic moments of fluctuations. The stronger singular
behavior of higher moments of fluctuations makes them more attractive
signatures of the QCD critical point~\cite{Stephanov:2008qz}. A study
of the higher-order moments would require generalization of the
analysis to nonlinear equations such as
(\ref{eq:boltz-noncov}) and~(\ref{eq:sigma-f}).

\acknowledgments

The hospitality of the Institute for Nuclear Theory at the University
of Washington during the program ``The QCD Critical Point'', which
stimulated this project, is gratefully acknowledged. The author thanks
K.~Rajagopal and D.~Son for comments and discussion.
This work is supported by
the DOE grant No.\ DE-FG0201ER41195.

\appendix
\section{Notations}
\label{sec:notations}


\begin{list}{}{\itemsep 0pt \parsep 0pt  \itemindent -2em }
\item
  ${\cal C}[f]$ --  collision integral (\ref{eq:boltzmann-part});
\item
  $D$ -- diffusion constant (\ref{eq:D-D_0});
\item
  $D_0$ -- same, at $g=0$, (\ref{eq:D_0});
\item
  $\mathbb E$ -- ``entropy matrix''~(\ref{eq:ent-matrix}),~(\ref{eq:E-def});
\item
  $f$ -- short for $f(\bm x,\bm p;t)$, non-equilibrium distribution
  function (\ref{eq:boltzmann-part});
\item
  $f_\sigma$ -- equilibrium distribution function for given
    background $\sigma$~(\ref{eq:f-sigma});
\item
  $f_0$ or $f_0'$ -- short for $f_0(\bm p)$ or $f_0(\bm p')$, as above, for $\sigma=0$;
\item
  $\mathbb G$ -- operator (\ref{eq:G-def}), acting on $\bm h$ gives ``drift''
    terms in stochastic
    equations~(\ref{eq:linear-noise}),~(\ref{eq:lin-noise-G});
\item 
    $g$ -- coupling of particles to $\sigma$ (\ref{eq:g-def});
\item 
    $h$ -- short for $h(\bm x,\bm p; t)$
    relative deviation of $f$ from $f_\sigma$~(\ref{eq:f-h}) ;
\item 
    $\hmean$ -- mean value of $h$~(\ref{eq:h-tilde-h-0});
\item 
    $\tilde h$ -- deviation of $h$ from $\hmean$~(\ref{eq:h-tilde-h-0});
\item 
    $\bm h$ -- generalized vector $(h,\pi, \sigma)$~(\ref{eq:ent-matrix});
\item 
    $\bm h_{\lambda}$ -- eigenmode of $\mathbb G$ with eigenvalue
    $\lambda$ (\ref{eq:lambda-h-G-h});
\item 
    $\hat{\bm h}$ -- convenient basis vector (\ref{h-hat});
\item 
    ${\cal I}$ -- linearized collision integral~(\ref{eq:collision-linearized});
\item 
    ${\cal K}$ -- linear operator related to ${\cal I}$ by Eq.~(\ref{eq:K});
\item 
    $M$ or  $M(\sigma)$ -- particle mass for given $\sigma$ (\ref{eq:action});
\item 
    ${\cal M}[h]$ -- mean value of $h$ 
    (\ref{eq:M-def});
\item 
    $m_0$ -- vacuum mass of the field
    $\sigma$~(\ref{eq:sigma-linearized}), $d^2U(0)/d\sigma^2$;
\item 
    $m$  -- thermal screening mass of $\sigma$~(\ref{eq:m2-def}),~(\ref{eq:m-m_0});
\item 
    $\tilde m$ -- thermal pole mass (rest energy) of $\sigma$
    quasiparticles~(\ref{eq:m-tilde-def}),~(\ref{eq:lambda-1-2});
\item 
    $\mc$ or $\mk$ -- screening masses of $\sigma$ at chemical or kinetic
    freezeout;
\item 
    $N$ -- total number of particles;
\item 
    $n_0$ -- equilibrium density of particles~(\ref{eq:n0});
\item 
    $\bm p$ or $\bm p'$ -- particle momentum variable in $f$;
\item 
    $p_T$ -- the magnitude of the component of $\bm p$ transverse to beam axis;
\item 
    $\bm q$ -- Fourier conjugate to $\bm x$ in $h(\bm x,\bm p; t)$;
\item 
    $\mathbb Q$ -- matrix of noise correlators~(\ref{eq:xixi-Q});
\item 
    $r_m$ -- ``memory'' factor (\ref{eq:corr-factor});
\item 
    $S^{(2)}$ -- quadratic terms in the entropy (\ref{eq:S-quadratic});
\item 
    $T$ -- temperature of the external bath~(\ref{eq:entropy}); 
\item 
    $U$ -- $U(\sigma)$ potential for $\sigma$, (\ref{eq:action});
\item 
    $V$ -- $\intx$, 3-volume;
\item 
    $\mathbb V$ -- evolution operator (\ref{eq:dot-V});
\item 
    $\bm v$ -- short for $\bm v(\bm p)$, particle velocity (\ref{eq:v-gamma});
\item 
    $\Gamma_0$ -- relaxation rate of $\sigma$ due to interaction with
    the external thermal bath only~(\ref{eq:lin-noise-pi});
\item 
    $\Gamma$ -- the full relaxation rate of $\sigma$ (\ref{eq:lambda-2-quadratic}),~(\ref{eq:Gamma-tilde-def});
\item 
    $\gamma$ or $\gamma'$ -- $\gamma(\bm p)$ or $\gamma(\bm p')$, relativistic
    factor~(\ref{eq:v-gamma});
\item 
    $\deli^2$ -- Eq.~(\ref{eq:deli-def}) and also $\tilde
    m^2-m^2$~(\ref{eq:m-tilde-def});
\item 
    $\delta_{\bm p \bm p'}$ --  $(2\pi)^3\delta^3(\bm p -\bm p')$, (\ref{eq:nu-correlator-equilibrium});
\item 
    $\eta$ -- $\eta(x)$, Langevin noise in eq.~(\ref{eq:lin-noise-pi});
\item 
    $\bm\hdual_{\lambda}$ -- dual vector to $\bm
    h_\lambda$~(\ref{eq:bar-h-lambda});
\item 
    $\hdualhat$ -- see Eq.~(\ref{eq:h-dual-hat});
\item 
    $\lambda_0$ -- smallest eigenvalue of~(\ref{eq:lambda-h-G-h});
\item 
    $\nu_{\bm p}$ -- momentum space distribution, (\ref{eq:mom-density});
\item 
    $\xi$ -- $\xi(\bm x,\bm p; t)$, noise in Boltzmann eqn.~(\ref{eq:lin-noise-h});
\item 
    $\bm\xi$ -- noise vector $(\xi,\eta,0)$~(\ref{eq:lin-noise-G}), (\ref{eq:xixi-Q});
\item 
    $\pi$ -- canonical momentum for $\sigma$, (\ref{eq:sigma-pi});
\item 
    $\bm\Sigma$ -- matrix of correlators $\langle\bm h\otimes\bm
    h\rangle$, (\ref{eq:hh-Sigma});
\item 
    $\tilde{\bm\Sigma}$ -- deviation of $\bm\Sigma$ from equilibrium
    (\ref{eq:dot-tilde-Sigma});
\item 
    $\sigma$ -- $\sigma(x)$, scalar field, critical mode;
\item 
    $\hat{\bm\sigma}$ -- basis vector~(\ref{s-hat});
\item 
    $\tau$ -- collisional relaxation time (\ref{eq:I-tau});
\item 
    $\tau_e$ -- fireball evolution time scale, Sec.~\ref{sec:prelim};
\item 
    $\psi$ -- $\psi(\bm v^2)$, solution to Eq.~(\ref{eq:psi-def});
\item 
    $\int_{\bm p}$ -- see Eq.~(\ref{eq:int-p});
\item
  $\bm\cdot$ -- the scalar product is defined in Eq.~(\ref{eq:ycdot}).

\end{list}

\bibliography{bl}

\end{document}